# Mono-drive single-sideband modulation via optical delay lines on thin-film lithium niobate

Yikun Chen[1,2], Hanke Feng[1,2], Zhenzheng Wang[1,2], Ke Zhang[1,2], Xiangzhi Xie[1,2], Yuansong Zeng[1,2], Yujie Ren[1], Cheng Wang[1,2,*]

1 Department of Electrical Engineering, City University of Hong Kong, Kowloon, Hong Kong, China
2 State Key Laboratory of Terahertz and Millimeter Waves, City University of Hong Kong, Kowloon, Hong Kong, China
*cwang257@cityu.edu.hk

**Abstract**

Optical single-sideband (SSB) modulation features high spectral efficiency, substantial dispersion tolerance, and straightforward detection, making it a versatile technology for applications in optical communications, microwave photonics, optical sensing, satellite communication, etc. However, conventional SSB generators typically require two radio-frequency (RF) signals with a 90° phase difference to drive a pair of parallel phase or amplitude modulators, resulting in high system complexity and low power efficiency. In this paper, we propose and realize a simplified SSB generation scheme necessitating only a single RF drive, by achieving effective RF phase shift using on-chip optical delay lines. This approach not only reduces system complexity and saves energy consumption by 3 dB, but also enables easy scalability to higher frequencies. We demonstrate both full-carrier SSB (FC-SSB) and carrier-suppressed SSB (CS-SSB) modulation on thin-film lithium niobate platform. For FC-SSB, we show a maximum sideband suppression of 22.1 dB at 50 GHz and apply it to address the frequency-selective power fading problem in optical communication systems. For CS-SSB, we show a maximum sideband suppression of 22.5 dB and a sideband-to-carrier suppression of 16.9 dB at 50 GHz, which can act as an optical frequency shifter by sweeping the modulation frequencies. Moreover, the shifted optical frequency can be transferred back to the electrical domain by beating with a reference signal generated via a phase modulator on the same chip, achieving broadband RF frequency shifting from a maximum of 50 GHz down to 1 GHz. Our simple, power-efficient, and low-cost SSB modulation scheme could provide an effective solution for future high-frequency direct detection-based communication systems, frequency-modulated continuous wave radar/LiDAR, optical vector network analyzers, and microwave photonics systems.

## Introduction

The ever-growing data transmission demand for applications such as 5G/6G networks, Internet of Things (IoT), Massive Multiple Input Multiple Output (MIMO) and Radio-over-Fiber (RoF) systems imposes significant pressure on base stations and data center infrastructure, where optical communication systems with high efficiency, low cost, less complexity are strongly demanded [1]. Direct detection systems that utilize a single photodetector for signal reception offer the advantages of low cost and system simplicity, in contrast to coherent detection systems that require digital signal processors (DSP) and local oscillators [2]. However, the standard double-sideband (DSB) modulated signal in direct detection systems suffers from limited transmission distance due to frequency-selective power fading caused by chromatic dispersion [3]. In comparison, single-sideband (SSB) signal not only addresses this issue but also provides doubled spectral efficiency, making it widely used in intensity modulation/direct detection (IM/DD) systems [4–6]. Recently, SSB has also proven compatible with coherent systems in Kramers-Kronig receivers [7,8], enabling the reception of complex modulation format signals with further increased data transmission rates. Beyond optical communications, SSB modulators have also found applications in optical vector network analyzers [9,10], cold atom interferometry [11,12], frequency-modulated continuous-wave (FMCW) light detection and ranging (LiDAR) [13,14], microwave photonics [15,16], and quantum information systems [17–20].

SSB signals can be generated either by acousto-optic (AO) [21–23] or electro-optic (EO) modulation [24–33]. AO modulators offer a high sideband suppression ratio and are commonly used as frequency shifters [34,35]. However, they typically operate at lower modulation frequencies (usually in the MHz to a few GHz range) and often require a suspended structure to facilitate efficient phonon-photon interaction [36]. In contrast, EO modulators provide larger modulation bandwidths, making them more suitable for high-speed applications like data transmission, frequency-modulated continuous wave radar/LiDAR, and microwave photonics systems. Traditional EO modulators based on bulk lithium niobate platform have a bandwidth of ~35 GHz mainly limited by the weak optical mode confinement [37]. This constraint has been transcended by the recently emerged thin-film lithium niobate (TFLN) platform, which features much better optical confinement and higher EO modulation efficiency than its traditional counterpart, while maintaining the large transparent window and low optical loss of lithium niobate material [38,39]. TFLN EO modulators with ultrabroad bandwidths exceeding 100 GHz have been successfully demonstrated [40,41]. However, different from AO modulation, where the carrier energy can be unidirectionally transferred either to upper sideband (Stokes) or lower sideband (anti-Stokes) by designing the wave-vector direction of the acoustic wave, EO modulation naturally generates sidebands on both sides of the carrier, leading to DSB signals. To suppress one of the sidebands, various optical filters based on ring resonator [42], Bragg grating [43], and ring-assisted Mach-Zehnder interferometer (MZI)[10] have been adopted, where resonant structures with high quality factors are typically required to achieve sharp filter edges, making them sensitive to environment fluctuations. Alternatively, SSB modulation can be achieved through the destructive interference of two modulated optical paths that carry opposite phases in the two sidebands, which is the standard method employed in commercial SSB modulators. To achieve the necessary phase difference between sidebands, a pair of parallel phase modulators [in the case of dual-drive Mach-Zehnder modulators (DDMZM), for full-carrier SSB (FC-SSB)] [24–27] or amplitude modulators [in the case of in-phase/quadrature (IQ) modulators, for carrier-suppressed SSB (CS-SSB)] [12–15,28–33] driven by two radio-frequency (RF) signals are typically needed. Specifically, the two RF signals are required to have a 90° phase difference to form a Hilbert transform pair, which can be generated by two synchronized RF sources [e.g. dual-channel arbitrary waveform generator (AWG)] [13,25] or by splitting one RF source with a 90° RF hybrid [14,15,27,30,31,33]. The required additional discrete RF equipment not only is bulky and costly especially at high frequencies (commercial 90° RF hybrid is limited to < 40 GHz), but also introduces additional RF loss. The complex configuration of the parallel modulation pairs effectively incurs a 3 dB attenuation in modulation efficiency as the total RF power is equally distributed to two parallel electrodes. Additionally, as each electrode is typically centimeters long to achieve low half-wave voltage, this parallel driving method occupies a considerably large footprint to accommodate the redundant electrodes.

In this paper, we propose a simplified mono-drive SSB modulation scheme on TFLN that necessitates only a single RF drive for both FC-SSB and CS-SSB modulation, by performing the RF phase shifting function in the optical domain through on-chip optical delay lines. This design not only reduces the device footprint by half, but more importantly removes one set of redundant RF source that saves half of the modulation energy. For FC-SSB, we show a maximum sideband suppression of 22.1 dB and prove that the generated SSB signal has good resistance to the frequency-selective power fading problem. For CS-SSB, we attain a maximum sideband suppression of 22.5 dB and a sideband-to-carrier suppression of 16.9 dB, which could act as an optical frequency shifter by sweeping the modulation frequencies. By beating the shifted optical sideband with a nearby reference signal generated through a phase modulator on the same chip, we further demonstrate RF frequency conversion from 50 GHz down to 1 GHz. Moreover, different from RF 90° hybrid or dual-channel AWG that become increasingly more costly and lossy at higher frequencies, our photonic RF phase shifter can be easily scaled to higher frequency bands by appropriately designing the optical delay line length without incurring significantly higher loss, leveraging the excellent scalability and low loss of our TFLN platform. Our proposed schemes offer significant advantages in high-frequency SSB modulation and could provide low-cost, low-power SSB signal generation for future direct detection communication systems, FMCW radar/LiDAR, optical vector network analyzers, microwave photonics, and cold atom interferometry systems.

**Results**

**Working principle and device design**

Figure 1 presents a schematic comparison between conventional dual-drive SSB generators and our simplified approaches that employ only one RF signal. Figure 1 (a) shows the schematic diagrams of typical FC-SSB modulation based on DDMZM and

CS-SSB modulation based on IQ modulator. The two sets of parallel phase or amplitude modulators are driven by two RF signals with a 90° phase difference. In contrast, our mono-drive SSB modulation scheme achieves photonics-based RF phase shifters using on-chip optical delay lines (illustrated as spirals). Specifically, for FC-SSB generation [Fig. 1 (b), top], the input light is split into two branches and modulated by a single-drive RF signal, similar to that in a conventional push-pull DSB modulation. The amplitudes and phases of the modulated signals in these two branches are indicated in the green insets (i and iv). To generate the effective 90° RF phase shift, the modulated optical signal in the top branch (i) then passes through an optical delay line with a delay time equivalent to one-quarter of the period ($\tau_m$) of the target RF frequency. This introduces additional relative phases between carrier and the two sidebands ($\pi/2$ for upper sideband and $-\pi/2$ for lower sideband) that aligns the phases of the two sidebands (ii). The delay time is controlled by a carefully designed delay length (see Supplement 1 for details), where the group index of the delay-line waveguide is simulated using a commercial Finite Difference Eigenmode (FDE) solver (Ansys Lumerical, Mode Solutions). This benefits from the ability to route optical signals with small bending radii and the high-precision nano-fabrication process in the integrated platform, which would otherwise be highly challenging in the conventional weakly-confined waveguide platform. An additional optical phase shift (iii) is then induced by a thermal phase shifter to fine-tune the phase difference and selectively suppress one sideband via destructive interference between the two branches (iii and iv) when they recombine (see detailed theoretical analysis in Supplement 1).

For CS-SSB generation, the initially modulated light is further split into four branches [labeled a,b,c,d in Fig. 1 (b) bottom]. The bottom insets I and II indicate the amplitudes and phases of the modulated signals in the top (a, b) and bottom (c, d) two branches, respectively. A $\pi$ phase shift between top and bottom branches is introduced by applying a DC voltage onto the electrodes to adjust the bias point, which can alternatively be replaced by an additional thermal phase shifter before they finally recombine. Optical delay lines are then introduced in paths b and c to induce the 90° photonic RF phase shift, with delay lengths following the same principle as in the FC-SSB case. The four branches recombine twice for suppressing sidebands and carriers respectively. The recombination topology can be designed to either suppress the sidebands first and then the carrier, or vice versa. Here, to align with the FC-SSB design, the sidebands are first canceled by first combining branches a and c (leading to spectrum shown in inset V) as well as b and d (inset VI) with the assistance of a waveguide crossing [designed through full 3D finite-difference time-domain (FDTD) simulations (Ansys Lumerical, FDTD Solutions)] that we have characterized in previous work [44]. Two thermal phase shifters are utilized to ensure the correct relative phase for destructive interference and sideband selection (III and IV). Finally, the optical carrier is suppressed by combining the two output branches (V and VI) to realize CS-SSB modulation.

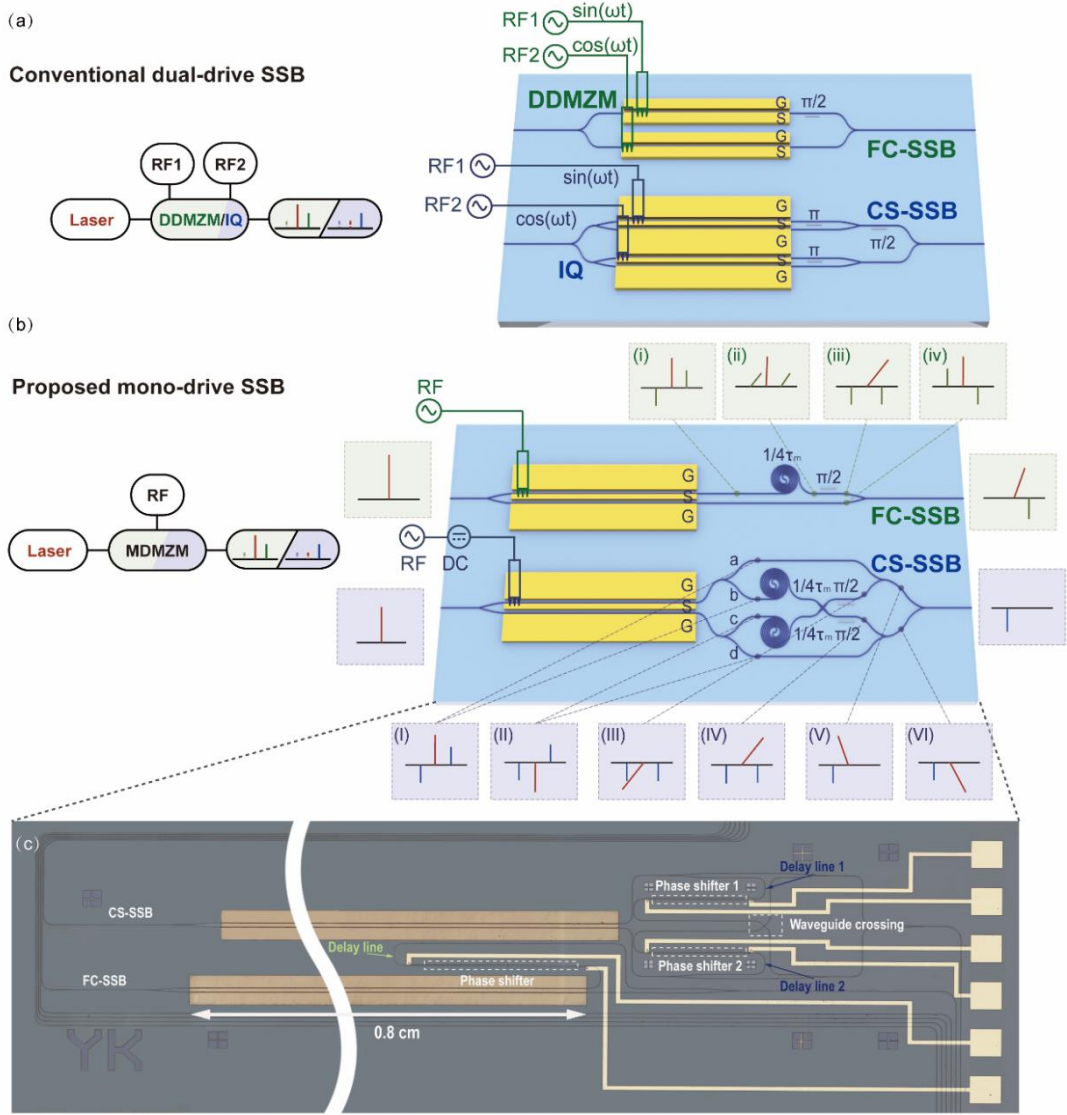

**Fig. 1.** (a) Schematic diagrams of conventional FC-SSB and CS-SSB generation using DDMZM and IQ modulators, respectively, driven by two separate RF signals. (b) Schematic diagrams of proposed simplified FC-SSB (top right) and CS-SSB (bottom right) generation using MDMZM with photonic RF phase shifters realized by optical delay lines. Insets (i-iv) and (I-VI) illustrate the amplitudes and phases of the optical carrier and sidebands at different locations of the FC-SSB and CS-SSB devices. (c) Microscope image of the fabricated devices. RF: radio-frequency, DDMZM: dual-drive Mach-Zehnder modulator, FC-SSB: full-carrier single-sideband, CS-SSB: carrier-suppressed single-sideband, MDMZM: mono-drive Mach-Zehnder modulator.

Figure 1 (c) shows the microscope image of the fabricated devices. We fabricate from a commercially available 4-inch x-cut TFLN wafer (NONOLN) consisting of a 500 nm LN thin film and a 4.7 μm buried $SiO_2$ bottom cladding on a 500 μm silicon substrate. $SiO_2$ is first deposited on the LN layer as a hard mask using plasma-enhanced chemical vapor deposition (PECVD). Micro-structures are then patterned on the entire wafer using an ASML UV Stepper lithography system die by die (1.5 cm × 1.5 cm) with a resolution of 500 nm. Next, the exposed resist patterns are transferred to the LN layer using an inductively coupled plasma reactive-ion etching (ICP-RIE) system, leading to a 250 nm LN rib waveguide and a 250 nm LN slab. Modulation electrodes with 500 nm copper and 50 nm gold are then fabricated using a sequence of photolithography, thermal evaporation, and lift-off process. Following this, a third repetitive process is used to fabricate the thermal heater with 100 nm nichrome. Finally, a fourth repetitive process is used for the wires/pads with 500 nm copper and 100 nm gold. Aluminum wire bonding is used to connect the bonding pads on the chip and a printed circuit board (PCB) for controlling the multi-port thermal phase shifters.

Compared to conventional dual-drive SSB schemes, the proposed approach offers an economical design that saves half the RF energy required to achieve the same modulation strength by removing one set of RF drive and reducing the on-chip space occupied by redundant signal electrodes.

**FC-SSB modulation**

We first characterize the sideband suppression performance of the FC-SSB modulator. A continuous wave pump light from a tunable telecom laser (Santec TSL-510) is first sent to a polarization controller to ensure transverse electric (TE) polarization and then edge-coupled to the LN chip using a lensed fiber. A single continuous RF signal is generated by a vector network analyzer (VNA, Keysight E5080B), amplified by an electrical power amplifier (SHF S807), and applied to the electrode using a GSG probe (GGB industries). DC power supplies are connected to a PCB for thermal phase shifter control. The modulated light is collected by another lensed fiber and monitored by an optical spectrum analyzer (OSA, Yokogawa AQ6370).

Figure 2 (a) shows the measured linear transmission of the FC-SSB modulator with a designed delay time of ~ 10 ps (corresponding to photonics RF 90° phase shift at 25 GHz) under various optical phase shifts. The corresponding modulated spectra at different bias points are shown in Fig. 2 (b)-(d). Lower sideband suppression of 17.6 dB and upper sideband suppression of 18 dB are measured respectively at the two quadrature points at 25 GHz. The highest sideband suppression is measured to be 20.6 dB at 27 GHz rather than the intended 25 GHz, which results from deviation between the group indices of the simulated and the first batch of fabricated devices (e.g. deviation in waveguide width and etching depth in fabrication). This deviation can be optimized by calibrating the fabrication error across different waveguides. The maximum sideband suppression in our current design architecture is constrained by the extinction ratio of the delay-line-inserted MZI, which could be affected by the loss imbalance between the two branches as they propagate different distances, as well as imperfection in the 50/50 waveguide splitters. The suppression ratio can be improved by optimizing the waveguide loss through improved design and fabrication. For example, adopting wider waveguides (e.g., increasing from 1.2 μm to 2 μm) can reduce propagation loss by reducing optical field overlap with surface roughness. This problem will also be alleviated in devices targeting higher frequencies with shorter delays, as shown in Fig. 2 (f)-(g), which showcases the device with a designed delay of 5 ps (corresponding to a target frequency of 50 GHz). The maximum measured sideband suppression is 22.1 dB, showing ~1.5 dB improvement. Similar to the traditional RF branch-line 90° hybrid coupler, which has a relative bandwidth of around 10% (within ±5° phase error) [45], our design also exhibits a working bandwidth limited by the walk-off of the photonic-induced RF phase, as the induced phase changes with frequency for a fixed delay line. Here, we define the working bandwidth of our SSB modulators as the range over which sideband suppression decreases by 3 dB. Figure 2 (h) presents simulation results showing the relationship between sideband suppression and modulation frequency for devices with designed delay lines for 25 GHz (green), 50 GHz (blue), and 100 GHz (purple). It can be seen that each of the devices can provide SSB operation not only at the target frequency, but also at higher frequencies that occur periodically with a period twice the target frequency. These operation points correspond to cases where the delay time equals $(2n + 1)\tau_m/4$, where n can be any positive integer number. The enlarged view in Fig. 2 (i) illustrates the simulated 3 dB suppression bandwidths of those three devices, respectively, as 6.25 GHz, 12.5 GHz, and 25 GHz, corresponding to 25% of the target frequency. This is equivalent to a 11% relative bandwidth if we adopt the ±5° RF phase deviation caliber often used to characterize traditional RF branch-line 90° hybrid couplers, indicating a comparable bandwidth performance. Figure 2 (j) shows that the measured suppression ratios (green dots) at various frequencies align well with the predicted values (light green curve), with a measured 3 dB bandwidth of 6.75 GHz centered at 27 GHz. The sideband suppression of the simulation predicted curve has an upper limit calibrated by the extinction ratio of the MZI.

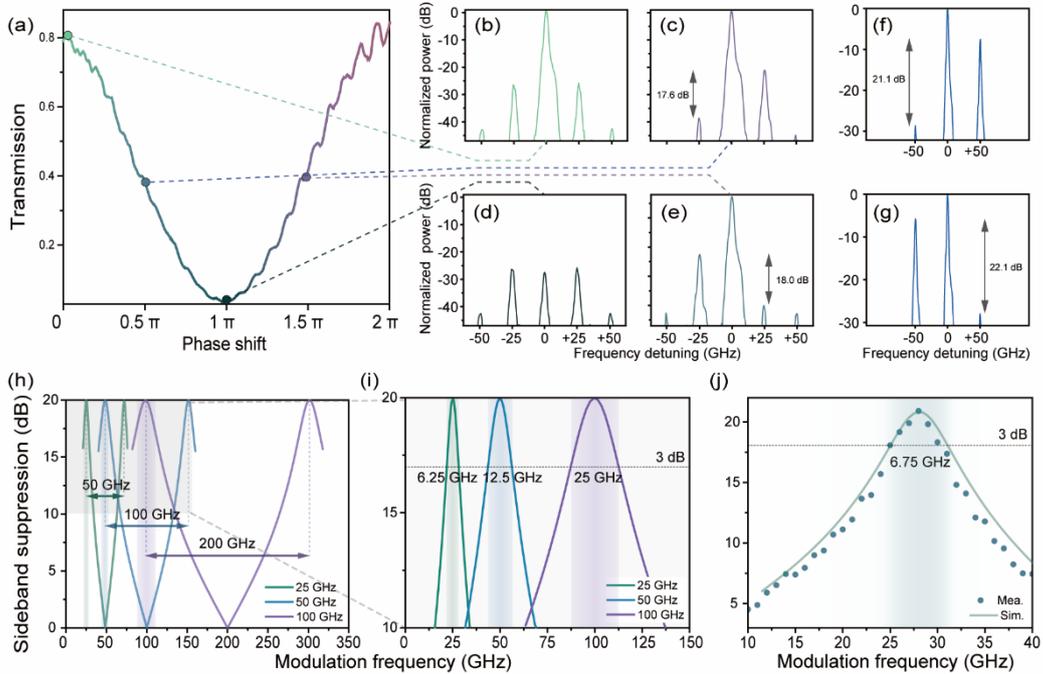

**Fig. 2.** (a) Normalized optical transmission of FC-SSB modulator at various phase shift. (b-e) Measured SSB spectra of device targeting 25 GHz biased at full point (b), null point (d), and the two quadrature points (c,e), respectively. (f-g) Measured SSB spectra of device targeting 50 GHz biased at quadrature points. (h-i) Simulated sideband suppression ratios as functions of modulation frequency for devices targeting 25 GHz (green), 50 GHz (blue), and 100 GHz (purple), respectively.

(j) Measured (dots) and simulated (line) sideband suppression ratios near the center frequency of 27 GHz.

We then verify that our proposed device could generate FC-SSB signals with strong resistance to the dispersion-selective frequency fading problem commonly encountered in DSB modulated signals. Figure 3 (a) shows the setup of the measurement system. Scanning RF signals (1~40 GHz) are generated from port 1 of a VNA and used to drive the SSB modulator. A DC power supply controls the bias point via the thermal phase shifters, enabling DSB (full point) or SSB (quadrature point) modulation. The modulated light is then sent through a 5 km dispersive single-mode fiber (G.652.D). The light is subsequently amplified by an erbium-doped fiber amplifier (EDFA, Amonics) and detected by a high-speed photodetector (PD, Finisar XPDV21X0RA, 50 GHz). The electrical signal from the PD is finally sent back to port 2 of the VNA to measure the EO $S_{21}$ parameters.

The EO $S_{21}$ of DSB (orange curve) and SSB signals (blue curve) are shown in Fig. 3 (b). It can be seen that the DSB signal degrades severely at ~26 GHz due to the dispersion-induced destructive interference between the upper and lower sidebands at the PD. In contrast, our generated SSB signal shows strong resistance to this power fading issue with a flat spectral response and an improvement of over 30 dB in signal power near 26 GHz.

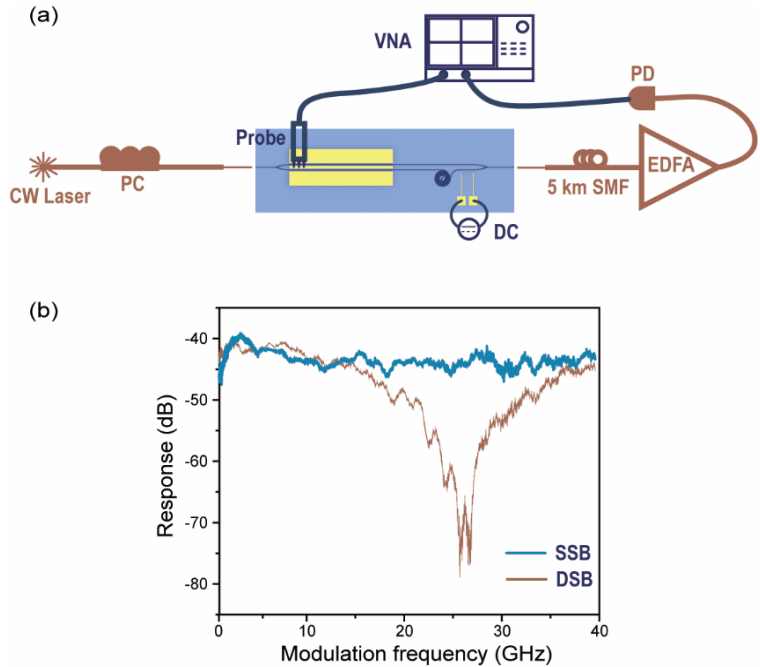

**Fig. 3.** (a) Schematic diagram of the frequency-selective fading measurement setup. (b) Measured EO $S_{21}$ of the DSB (orange) and FC-SSB (blue) signals after transmitting through 5 km fiber. CW: continuous wave, PC: polarization controller, VNA: vector network analyzer, EDFA: erbium-doped fiber amplifier, PD: photodetector, SSB: single-sideband, DSB: double sideband.

**CS-SSB Modulation**

The characterization of the CS-SSB modulator is presented in Figure 4. The measurement setup illustrated in Fig. 4(a) is similar to that used for FC-SSB characterization, incorporating an additional DC power supply and a bias-T for extra phase control. The measured SSB spectrum of the device centered at 25 GHz is shown in Fig. 4 (b), where a sideband suppression ratio of 16.6 dB and a sideband-to-carrier suppression ratio of 16.8 dB are achieved. Compared to the original carrier (dashed line), a carrier suppression of 23.1 dB is attained, with 22.3 dB resulting from destructive interference in our device. The rest 0.7 dB suppression comes from the inherent carrier suppression characteristics of the EO modulation process, which follows the Bessel function of its first kind under a modulation power of approximately 15 dBm.

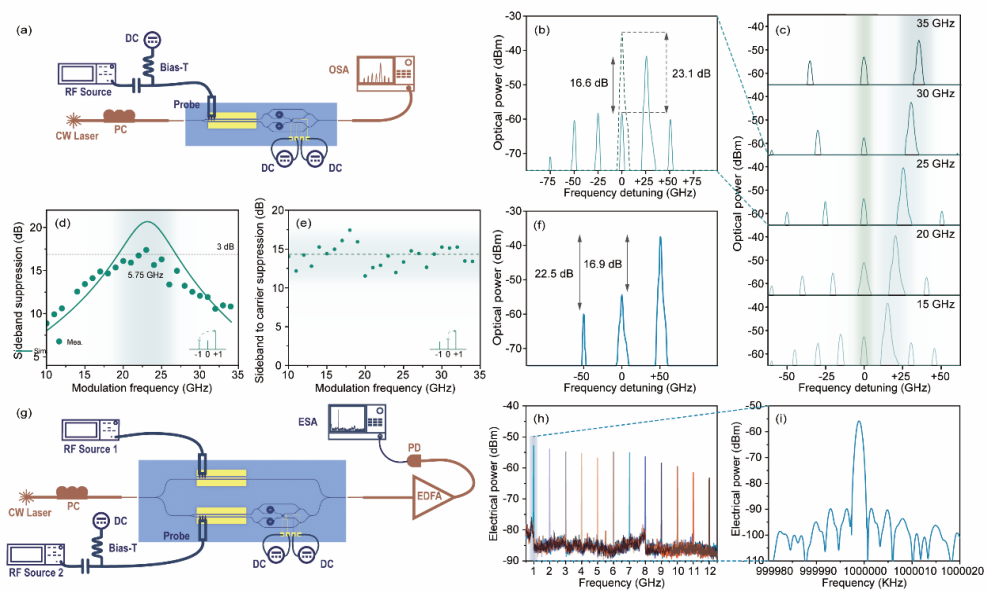

**Fig. 4.** (a) Schematic diagram of the CS-SSB measurement setup. (b) Measured output spectrum of the device under 25 GHz modulation. Dashed curve shows the input optical carrier. (c) Optical spectra of frequency shifted signals at various RF frequencies. (d) Measured (dots) and simulated (line) sideband suppression of the 25 GHz device at different modulation frequencies. (e) Measured sideband-to-carrier suppression across various modulation frequencies. (f) Measured CS-SSB spectra of the device centered at 50 GHz. (g) Schematic diagram of the setup for RF frequency shifting. The second-order sideband generated by the phase modulator (top path) is used as the reference signal to beat with the CS-SSB signal (bottom path). (h) Frequency-shifted RF signals by beating various CS-SSB signals that are 39-50 GHz away from optical carrier with a fixed reference signal (51 GHz from carrier) generated by the phase modulator in (g). (i) Enlarged view of the RF signal in (h) that is shifted to 1 GHz. OSA: optical spectrum analyzer, ESA: electrical spectrum analyzer.

As the RF frequency changes, the CS-SSB optical frequency shifts accordingly, acting as an optical frequency shifter, as shown in Fig. 4 (c). Detailed sideband suppression and sideband-to-carrier suppression at various modulation frequencies are depicted in Figs. 4 (d) and 4 (e). The measured 3 dB operation bandwidth is 5.75 GHz, centered in this case at 23 GHz. The measured sideband suppression (dots) is not as high as the simulation-predicted value (curve) near the center frequency, likely due to non-uniform lithium niobate film thickness and etching depth, which introduce additional loss and delay imbalances between the upper and lower branches. Figure 4 (e) shows the measured sideband-to-carrier suppression ratios at various modulation frequencies, with an average suppression of around 15 dB. The fluctuations at different frequencies likely arise from phase deviations in the two thermal phase shifters during measurement. The output spectrum of another device centered at 50 GHz is shown in Fig. 4 (f), where a sideband suppression of 16.9 dB and a sideband-to-carrier suppression of 22.5 dB are achieved under a modulation power of approximately 10 dBm. This frequency-shifted signal can be converted back to the electrical domain by beating it with a reference optical signal, effectively serving as an RF frequency shifter, as depicted in Fig. 4(g). The reference signal is generated by a phase modulator fabricated in parallel with the CS-SSB modulator on the same TFLN chip, where a fixed 25.5 GHz RF modulation signal (generated by Anritsu MG369C) is used to generate second-order sidebands as a reference near the CS-SSB signal. The sharing of optical carrier between the phase and CS-SSB modulators ensures a low-noise photonic mixing process. The swept optical CS-SSB signal (39-50 GHz) and the fixed second-order sidebands (51 GHz from the carrier) of the phase modulator beat at the PD (Newport 1544A, 12 GHz), which is analyzed by an electrical spectrum analyzer (ESA, Rohde Schwarz FSW43). Shifted signals spanning a broad frequency range from 1 GHz to 12 GHz are shown in Fig. 4 (h). Figure 4 (i) details the measured RF signal with a maximum frequency shift down to 1 GHz, indicating a narrow linewidth of 1.0 kHz.

**Conclusion**

In this paper, we have proposed and demonstrated delay-line assisted mono-drive schemes for power-efficient and compact FC-SSB and CS-SSB generations. The adopted delay lines serve as photonics RF 90° phase shifters, effectively simplifying the electrical configuration and saving 50% of the electrical power by removing one set of redundant electrodes, which provides economical solutions for compact SSB modulators. A maximum sideband suppression of 22.1 dB and 22.5 dB are demonstrated for 50 GHz FC-SSB and CS-SSB devices, along with a 16.9 dB sideband-to-carrier suppression for the CS-SSB. Meanwhile, our proposed schemes offer a 3dB sideband suppression bandwidth of 25% of the center frequency, which is comparable to the traditional schemes using branch-line 90° RF hybrid couplers. We further showcase that the FC-SSB device could provide modulated optical signals with strong resistance to the frequency-selective power fading problem in optical communications, while the CS-SSB scheme functions as both an optical frequency shifter by varying the modulation frequency, and an RF frequency shifter by beating the shifted optical signal with a reference optical signal generated on the same chip. Importantly, unlike electronics-based RF 90° hybrid or dual-channel AWG, our photonic RF phase shifter can be easily scaled to higher frequencies by appropriately designing the optical delay line length without compromise in loss and phase-shift precision, which has significant advantages for high-frequency SSB generations. The proposed delay-line assisted SSB generation scheme is compatible with various integrated platforms and has broad applications in optical communications, microwave photonics, frequency-modulated LiDAR/radar systems, optical vector network analyzers, optical sensing, and cold atom interferometry systems.

**Funding.** Research Grants Council, University Grants Committee (N_CityU113/20, CityU 11212721); Croucher Foundation (9509005); City University of Hong Kong (9610682).

**Acknowledgments.** We acknowledge Nanosystem Fabrication Facility (CWB) of HKUST for the device/system fabrication.

**Disclosures.** The authors declare no conflicts of interest.

**Data availability.** Data underlying the results presented in this paper are not publicly available at this time but may be obtained from the authors upon reasonable request.

**Supplemental document.** See Supplement 1 for supporting content.

# References

1. J. Krause Perin, A. Shastri, and J. M. Kahn, "Data center links beyond 100 Gbit/s per wavelength," Optical Fiber Technology **44**, 69–85 (2018).
2. M. Chagnon, "Optical communications for short reach," J. Lightwave Technol. **37**, 1779–1797 (2019).
3. A. S. Meijer, G. Berden, D. D. Arslanov, *et al.*, "An ultrawide-bandwidth single-sideband modulator for terahertz frequencies," Nature Photon **10**(11), 740–744 (2016).
4. T. Bo, H. Kim, Z. Tan, *et al.*, "Optical single-sideband transmitters," J. Lightwave Technol. **41**, 1163–1174 (2023).
5. Y. Wang, H. Li, M. Cheng, *et al.*, "Experimental demonstration of secure 100 Gb/s IMDD transmission over a 50 km SSMF using a quantum noise stream cipher and optical coarse-to-fine modulation," Opt. Express **29**(4), 5475 (2021).
6. B. Baeuerle, C. Hoessbacher, W. Heni, *et al.*, "100 GBd IM/DD transmission over 14 km SMF in the C-band enabled by a plasmonic SSB MZM," Opt. Express **28**(6), 8601 (2020).
7. A. Mecozzi, C. Antonelli, and M. Shtaif, "Kramers–Kronig coherent receiver," Optica **3**(11), 1220 (2016).
8. A. J. Lowery and T. Feleppa, "Analog low-latency Kramers-Kronig optical single-sideband receiver," J. Lightwave Technol. **39**, 3130–3136 (2021).
9. S. Pan and M. Xue, "Ultrahigh-resolution optical vector analysis based on optical single-sideband modulation," J. Lightwave Technol. **35**, 836–845 (2017).
10. H. Feng, T. Ge, Y. Hu, *et al.*, "In-situ optical vector analysis based on integrated lithium niobate single-sideband modulators," (n.d.).
11. J. Lee, R. Ding, J. Christensen, *et al.*, "A compact cold-atom interferometer with a high data-rate grating magneto-optical trap and a photonic-integrated-circuit-compatible laser system," Nat Commun **13**(1), 5131 (2022).
12. A. Kodigala, M. Gehl, G. W. Hoth, *et al.*, "High-performance silicon photonic single-sideband modulators for cold-atom interferometry," Sci. Adv. **10**(28), eade4454 (2024).
13. M. Kamata, Y. Hinakura, and T. Baba, "Carrier-suppressed single sideband signal for FMCW LiDAR using a Si photonic-crystal optical modulators," J. Lightwave Technol. **38**, 2315–2321 (2020).

14. P. Shi, L. Lu, C. Liu, *et al.*, "Optical FMCW signal generation using a silicon dual-parallel Mach-Zehnder modulator," IEEE Photon. Technol. Lett. **33**, 301–304 (2021).
15. F. Yang, D. Wang, Y. Wang, *et al*, "Photonics-assisted frequency up/down conversion with tunable OEO and phase shift," J. Lightwave Technol. **38**, 6446–6457 (2020).
16. Z. Tang, Y. Li, J. Yao, *et al.*, "Photonics-based microwave frequency mixing: methodology and applications," Laser & Photonics Reviews **14**, 1800350 (2020).
17. H.-P. Lo and H. Takesue, "Precise tuning of single-photon frequency using an optical single sideband modulator," Optica **4**(8), 919 (2017).
18. C. Chen, J. E. Heyes, J. H. Shapiro, *et al.*, "Single-photon frequency shifting with a quadrature phase-shift keying modulator," Sci Rep **11**(1), 300 (2021).
19. D. Milovancev, N. Vokic, F. Laudenbach, *et al.*, "High rate CV-QKD secured mobile WDM fronthaul for dense 5G radio networks," J. Lightwave Technol. **39**, 3445–3457 (2021).
20. F. Laudenbach, B. Schrenk, C. Pacher, *et al.*, "Pilot-assisted intradyne reception for high-speed continuous-variable quantum key distribution with true local oscillator," Quantum **3**, 193 (2019).
21. Z. Yu and X. Sun, "Gigahertz acousto-optic modulation and frequency shifting on etchless lithium niobate integrated platform," ACS Photonics **8**, 798–803 (2021).
22. E. A. Kittlaus, W. M. Jones, P. T. Rakich, *et al.*, "Electrically driven acousto-optics and broadband non-reciprocity in silicon photonics," Nat. Photonics **15**(1), 43–52 (2021).
23. L. Shao, N. Sinclair, J. Leatham, *et al.*, "Integrated microwave acousto-optic frequency shifter on thin-film lithium niobate," Opt. Express **28**(16), 23728 (2020).
24. X. Zhang, C. Zhang, C. Chen, *et al.*, "Digital chromatic dispersion pre-management for SSB modulation direct-detection optical transmission systems," Optics Communications **427**, 551–556 (2018).
25. X. Gao, B. Xu, Y. Cai, *et al.*, "QAM modulation with single DDMZM based on direct-detection and Kramers-Kronig scheme in long reach PON," Optical Fiber Technology **48**, 289–296 (2019).
26. Y. Cai, X. Gao, Y. Ling, *et al.*, "Performance comparison of optical single-sideband modulation in RoF link," Optics Communications **463**, 125409 (2020).
27. A. Loayssa, C. Lim, A. Nirmalathas, *et al.*, "Design and performance of the bidirectional optical single-sideband modulator," J. Lightwave Technol. **21**(4), 1071–1082 (2003).
28. F. Paloi and S. Haxha, "Analysis of the carrier suppressed single sideband modulation for long distance optical communication systems," Optik **161**, 230–243 (2018).
29. D. Fonseca, A. V. T. Cartaxo, and P. Monteiro, "Optical single-sideband transmitter for various electrical signaling formats," J. Lightwave Technol. **24**(5), 2059–2069 (2006).
30. B. Hraimel, X. Zhang, Y. Pei, *et al.*, "Optical single-sideband modulation with tunable optical carrier to sideband ratio in radio Over fiber systems," J. Lightwave Technol. **29**, 775–781 (2011).
31. T. Kawanishi and M. Izutsu, "Linear single-sideband modulation for high-SNR wavelength conversion," IEEE Photon. Technol. Lett. **16**, 1534–1536 (2004).
32. K. Higuma, S. Oikawa, Y. Hashimoto, *et al.*, "X-cut lithium niobate optical single-sideband modulator," Electron. Lett. **37**(8), 515 (2001).
33. X. Lei, G. Wang, H. Tan, *et al.*, "Optical carrier-suppressed single sideband modulation based on a thin-film lithium niobate IQ modulator for FMCW ranging application," J. Lightwave Technol. 1–7 (2024).
34. N. Li, C. P. Ho, S. Zhu, *et al.*, "Aluminium nitride integrated photonics: a review," Nanophotonics **10**(9), 2347–2387 (2021).

35. J.-H. Kim, S. Aghaeimeibodi, J. Carolan, *et al*., "Hybrid integration methods for on-chip quantum photonics," Optica **7**(4), 291 (2020).
36. P. Delsing, A. N. Cleland, M. J. A. Schuetz, *et al*., "The 2019 surface acoustic waves roadmap," J. Phys. D: Appl. Phys. **52**(35), 353001 (2019).
37. E. L. Wooten, K. M. Kissa, A. Yi-Yan, *et al*., "A review of lithium niobate modulators for fiber-optic communications systems," IEEE J. Select. Topics Quantum Electron. **6**(1), 69–82 (2000).
38. C. Wang, M. Zhang, X. Chen, *et al*., "Integrated lithium niobate electro-optic modulators operating at CMOS-compatible voltages," Nature **562**(7725), 101–104 (2018).
39. H. Feng, T. Ge, X. Guo, *et al*., "Integrated lithium niobate microwave photonic processing engine," Nature **627**(8002), 80–87 (2024).
40. F. Arab Juneghani, M. Gholipour Vazimali, J. Zhao, *et al*, "Thin-film lithium niobate optical modulators with an extrapolated bandwidth of 170 GHz," Advanced Photonics Research **4**, 2200216 (2023).
41. F. Valdez, V. Mere, and S. Mookherjea, "100 GHz bandwidth, 1 volt integrated electro-optic Mach–Zehnder modulator at near-IR wavelengths," Optica **10**(5), 578 (2023).
42. M. Tan, X. Xu, J. Wu, *et al*., "Orthogonally polarized RF optical single sideband generation with integrated ring resonators," J. Semicond. **42**(4), 041305 (2021).
43. J. Li, T. Ning, L. Pei, *et al*., "Performance analysis of an optical single sideband modulation approach with tunable optical carrier-to-sideband ratio," Optics & Laser Technology **48**, 210–215 (2013).
44. Y. Chen, K. Zhang, H. Feng, *et al*., "Design and resonator-assisted characterization of high-performance lithium niobate waveguide crossings," Opt. Lett. **48**(9), 2218 (2023).
45. A. A. Abdulbari, S. K. Abdul Rahim, P. J. Soh, *et al*., "A review of hybrid couplers: State-of-the-art, applications, design issues and challenges," Int J Numerical Modelling **34**, e2919 (2021).